\documentclass[conference,10pt,twocolumns]{IEEEtran}
\usepackage{epsf}
\usepackage{graphicx}
\usepackage{epsfig,latexsym,amsmath,epsf,amssymb,amsfonts}
\usepackage{amssymb}
\usepackage{placeins}
\usepackage{epsfig}
\usepackage{epstopdf}
\usepackage{color}
\usepackage[usenames,dvipsnames]{xcolor}

\begin{document}

\title{Effective Capacity of Delay Constrained Cognitive Radio Links Exploiting Primary Feedback }

\author{\large Ahmed H. Anwar$^\dagger$, Karim G. Seddik$^\ddagger$, Tamer ElBatt$^\dagger$ and Ahmed H. Zahran$^\dagger$ \\ [.1in]
\small  \begin{tabular}{c} $^\dagger$Wireless Intelligent
Networks Center (WINC), Nile University, Smart Village, Egypt.\\
$^\ddagger$Electronics Engineering Department, American University in Cairo, AUC Avenue, New Cairo 11835, Egypt.\\

email: ahmed.anwar@nileu.edu.eg, kseddik@aucegypt.edu, telbatt@ieee.org, ahzahran@ieee.org
\end{tabular} }

\maketitle

\begin{abstract}

In this paper, we analyze the performance of a secondary link in a cognitive radio (CR) system operating under statistical quality of service (QoS) delay constraints. In particular, we quantify analytically the performance improvement for the secondary user (SU) when applying a feedback based sensing scheme under the ``SINR Interference" model. We leverage the concept of effective capacity (EC) introduced earlier in the literature to quantify the wireless link performance under delay constraints, in an attempt to opportunistically support real-time applications. Towards this objective, we study a two-link network, a single secondary link and a primary network abstracted to a single primary link, with and without primary feedback exploitation. We analytically prove that exploiting primary feedback at the secondary transmitter improves the EC of the secondary user and decreases the secondary user average transmitted power. Finally, we present numerical results that support our analytical results.


\end{abstract}
\footnotetext[1]{This publication was made possible by NPRP 4-1034-2-385 and 
NPRP 09-1168-2-455 from the Qatar National Research Fund 
(a member of Qatar Foundation). The statements made herein 
are solely the responsibility of the authors.}
\footnotetext[2]{Tamer ElBatt and Ahmed H. Zahran are also affiliated with the EECE Dept., Faculty of Engineering, Cairo University.}

\section{Introduction}\label{Int}

Wireless communication is becoming more and more challenging as more users compete for limited bandwidth. Surprisingly, in some spectrum locations and at some times, $  70 \% $ of this reserved spectrum is idle \cite{federal2002spectrum}. Cognitive radios (CRs) have been studied extensively over the last decade due to its opportunistic, agile and efficient spectrum utilization merits. Those merits enable secondary users (SUs) to use the frequency bands allocated to the primary (licensed) users (PUs) without causing destructive interference to them. An overview of CR principals and challenges can be found in \cite{mitola1999cognitive,haykin2005cognitive,akyildiz2006next}. The coexistence of SUs and PUs is allowed provided that minimal, or no harm, is caused to the primary network.


In a typical CR setting, the secondary transmitter senses the primary user activity. The SU decides whether to access the channel or not according to the sensing outcome. This setting is problematic in the sense that cognitive users are not aware of their impact on the primary network, besides the usual sensing errors. In \cite{eswaran2007bits,huang2010distributed,lapiccirella2010cognitive}, the SU transmitter may exploit the information about the PU activity by overhearing the feedback  sent from the primary receiver to the primary transmitter before sensing the medium. For instance, optimizing the channel access decision has been proposed in \cite{seddikfeedback}, where the authors investigate the PUs stability and the SUs performance via exploiting the PU feedback under the collision model \cite{gupta2000capacity}.


Satisfying quality of service (QoS) requirements is another challenge for wireless networks, in particular, CR networks. Real-time applications, typically, require QoS guarantees, e.g., delay constraints. In order to incorporate a delay metric with the wireless link throughput, the notion of Effective Capacity is proposed in \cite{wu2003effective}. Effective Capacity (EC) is considered the wireless dual concept to the Effective Bandwidth which was originally coined for wired networks \cite{chang1995effective}.

The EC for interference and delay constrained CR relay channels is characterized in \cite{musavian2010effective} under Rayleigh fading channels.
In \cite{li2010effective}, the EC limits for CR networks, under peak interference constraints, are established. Moreover, it is shown that for a stricter delay requirement, the EC cannot benefit much from increasing the peak interference constraint. In addition, \cite{shakkottai2008effective} has studied a multi-user formulation of the EC with QoS constraints and proposed a channel-aware greedy scheduling policy as well as a channel-aware max-queue scheduling policy. It has been shown that those algorithms, which yield the same long-term throughput in the absence of QoS constraints, have drastically different performance when QoS constraints are imposed. However, none of the above studies has considered the effect of feedback on EC.

Our contribution in this paper is multi-fold. Mainly, We show that a higher EC for the CR link can be achieved by exploiting the PU receiver feedback messages. First, we explain and analyze the impact of overhearing the primary ARQ feedback on the EC of a CR link targeting opportunistic, real-time communications. Second, we extend the queuing-theoretic framework in \cite{seddikfeedback} to capture the role of primary feedback on the EC. Third, we conduct a mathematical analysis, based on the theory of non-negative matrices, for the signal to interference and noise ratio (SINR) model (aka the interference model) \cite{gupta2000capacity}. Finally, we prove that overhearing the primary receiver feedback not only improves the EC of the SU but also reduces the SU average transmitted power.

The rest of the paper is organized as follows. A background on the EC concept is given in Section \ref{EC}. The system model and underlying assumptions are presented in Section \ref{sysmod}. In Section \ref{Interf}, the EC problem with/without primary feedback exploitation is formulated and analyzed. Afterwards, numerical results and discussion are presented. Finally, conclusions and potential directions for future research are pointed out in Section \ref{Concl}.

\section{Background : Effective Capacity}\label{EC}
The conventional definition of information theoretic {\it capacity}, or {\it channel capacity}, is the tightest upper bound on the amount of information that can be reliably transmitted over a communications channel \cite{cover2006elements}. In \cite{wu2003effective}, Wu {\it et al.} introduced the fundamentally different notion of {\it effective capacity} (EC) of a general wireless link to be the maximum constant arrival rate that can be supported by a given channel service process while satisfying a statistical QoS requirement specified by the QoS exponent, denoted by $\theta$. The concept of EC is a link layer modelling abstraction to incorporate QoS 
requirements, e.g., delay, in performance analysis studies of wireless 
systems. Hence, we can capture the delay metric of a CRN without going into complex queuing analysis. 

If $Q$ is defined as the stationary queue length, then $\theta$ is the decay rate of the tail distribution of the queue length $Q$, that is
\begin{equation}
\lim_{q \to \infty }\frac{\log \Pr(Q\geq q)}{q} = -\theta.
\end{equation}

Practically, $\theta$, which depends on the statistical characterization of the arrival and service processes, establishes bounds on delay or buffer lengths, and targets values of the delay or buffer length violation probabilities. It has been established in \cite{wu2003effective} that the EC for a given QoS exponent $\theta$ is given by

\begin{equation}
-\lim_{t \to \infty }\frac{1}{\theta t}\log_{e}\mathbb{E}\left \{ e^{-\theta S(t)} \right \}=-\frac{\Lambda (-\theta)}{\theta},\label{eqn_EC}
\end{equation}
where $S(t)=\sum_{k=1}^t r(k)$ represents the time accumulated service process and $\lbrace r(k)$, $k$= $1, 2,...\rbrace$ is the discrete, stationary and ergodic stochastic service process.
\section{System Model}\label{sysmod}
\begin{figure}
\centering
\includegraphics[width=0.8\linewidth,height=0.15\textheight]{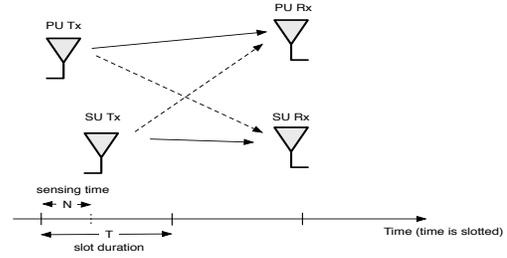}
\caption{System Model.}\label{fig_sysmod}
\vspace{-0.4 cm}
\end{figure}

In this paper, we consider a time slotted system as shown in Fig. \ref{fig_sysmod}. The primary network is abstracted by a primary link (i.e, our analysis is valid for any number of primary users). Having one frequency channel, the primary transmitter will access the channel whenever it has a packet to send in its queue. On the other hand, a single SU attempts to access the medium with a certain policy, to be described later, based on the spectrum sensing outcome. The SU is assumed to have a packet to send at the beginning of each time slot. Data is transmitted in frames of duration $T$ seconds, where each frame fits exactly in a single time slot. We assume that the first $N$ seconds of the frame duration $T$ are used by the SU to sense the licensed spectrum. Note that, although we assumed that we have one frequency channel, if we considered $ n $ channels, our analysis 
is still valid and applicable for each channel.

In the rest of this paper, the system that exploits the PU feedback messages is denoted by the ``feedback-aided system'', while the other system that does not exploit the PU feedback messages is denoted by the ``No feedback system''. The ECs of both systems are analyzed under the ``interference model'' \cite{gupta2000capacity}. The SU attempts to transmit packets with power $P_1$ ($P_2$) when the channel is sensed busy (idle), where $P_1 < P_2$. These power levels correspond to the SU transmission rates of $ r_1 $ and $ r_2 $ for busy and idle channels, respectively. Ideally the medium is sensed busy if the PU is sending a packet, however a misdetection occurs if such PU activity is not detected. On the other hand, the medium should be sensed idle if the PU is not sending any packets and false alarm occurs if the medium is sensed busy in this case. Simple energy detection \cite{akyildiz2006next} is adopted as the spectrum sensing mechanism.

The discrete time secondary link input-output relations for idle and busy channels in the $i^{th}$ symbol duration are given, respectively, by
\begin{equation}
y(i)=h(i)x(i)+n(i) \;\;\;\;\;\;\;\;   i=1,2, \cdots \label{relation1}
\end{equation}
\begin{equation}
y(i)=h(i)x(i)+s_p(i)+n(i) \;\;\;\;\;\;\;\;  i=1,2,\cdots, \label{relation2}
\end{equation}
where $x(i) $ and $ y(i)$ represent the complex-valued channel input and output, respectively. $h(i)$ denotes the fading coefficient between the cognitive transmitter and receiver, $s_p(i)$ is the interference coefficient from the primary network to the SU and $n(i)$ is the additive
thermal noise at the secondary receiver modeled as a zero-mean, circularly-symmetric complex Gaussian random variable with variance $\mathbb{E}{\lbrace|n(i)|^2}\rbrace$ = $ \sigma ^{2} _{n}$. The channel bandwidth is denoted by $B$. The channel input is subject to the following average energy constraints: $\mathbb{E}{\lbrace|x(i)|^2}\rbrace\leq P_1/B$ or $\mathbb{E}{\lbrace|x(i)|^2}\rbrace\leq P_2/B$ for all $i$'s, when the channel is sensed to be busy or idle, respectively. The fading coefficients are assumed to have arbitrary marginal distributions with finite variances, that is, $\mathbb{E}{\lbrace|h(i)|^2}\rbrace$ = $\mathbb{E}\lbrace z(i) \rbrace$ = $ \sigma ^{2}$ $<$ $\infty$, where $|h(i)|^2 = z(i)$. Finally, we consider a block-fading channel model and assume that the fading coefficients stay constant for a block of duration $T$ seconds (i.e., one frame duration) and change independently from one block to another.

In the proposed model, we leverage a perfect error-free primary feedback channel. The primary receiver sends a feedback at the end of each time slot to acknowledge the reception of packets. Typically, the PU receiver sends an ACK if a packet is correctly received, however, a NACK is sent if a packet is lost. Failure of reception is attributed to primary channel outage. In case of an idle slot, no feedback is sent. The SU is assumed to overhear and decode this primary feedback perfectly and to act as follows: if an ACK/no feedback is heard, the SU behaves normally and starts sensing the channel in the next time slot. However, if a NACK is overheard the SU transmits with a lower power in the next time slot. So that, ``sure'' high interference events are avoided since the reception of a NACK triggers the PU to retransmit in the next time slot with probability one.

In our model, we assume that the PU occupies the wireless channel with a fixed prior probability $\rho$ \cite{musavian2010effective}. The channel sensing can be formulated as a hypothesis testing problem between the additive white Gaussian noise $n(i)$ and the primary signal $s_p(i)$ in noise. Noting that there are $NB$ complex symbols in a duration of $N$ seconds, this can be expressed mathematically as follows:
\begin{equation}
H_0 : y(i)=n(i), \;\;\;\;\;\;\;\; i=1,...,N . B;
\end{equation}
\begin{equation}
H_1 : y(i)=s_p(i)+n(i), \;\;\;\;\;\;\;\; i=1,...,N . B.
\end{equation}

We can write down the probabilities of false alarm $P_f$ and detection $P_d$ as follows:
\begin{equation}\label{prob_f}
P_f=Pr(Y> \lambda | H_0)=1 - P\left ( \frac{NB\lambda}{\sigma ^{2} _{n} }  , NB \right);
\end{equation}
\begin{equation}\label{prob_d}
P_d=Pr(Y> \lambda | H_1) =1 - P\left ( \frac{NB\lambda}{\sigma ^{2} _{sp}+\sigma ^{2} _{n} }  , NB \right),
\end{equation}
where $ \lambda $ is the energy detector threshold, $Y = \frac{1}{NB} \sum_{i=1}^{NB} |y(i)|^2$ and $P(x, a)$ denotes the regularized lower gamma function defined as $P(x, a)=\frac{\gamma (x,a)}{\Gamma(a)}$ where $\gamma (x,a)$ is the lower incomplete gamma function. Note that the test statistic $ Y $ is chi-square distributed with $ 2NB $ degrees of freedom.

In the next section, the EC of both the feedback-aided system and the no feedback system is characterized and analyzed.

\section{EC Analysis under the SINR Interference Model}\label{Interf}
In this section, we characterize the EC of the SU for the no feedback and the feedback-aided systems. For the system with no feedback, we present the Markov chain modeling of the system and then calculate the EC of the SU. Then, the analysis is repeated for the feedback-aided system.

The interference model considered here can be thought of as a midway between the overlay model and the underlay model. It can also be thought of as a "hybrid", i.e., overlay/underlay \cite{akyildiz2006next}. The CR link inherits the channel sensing process from the overlay model, while co-existing with the PU, albeit at lower powers and rates, is inherited from the underlay model. The medium is busy with a constant probability $\rho$. Next we develop the underlying Markov chain model governing the system dynamics.

Generally, the sensing process outcome could be one of the following four possibilities:
\begin{enumerate}
\item Channel is busy and detected busy (correct detection), denoted (B-B): SU transmits using $ \lbrace r_1 ,P_1 \rbrace$, where $r_1$ is the transmission rate and $P_1$ is the transmission power;
\item Channel is busy and detected idle (misdetection), denoted (MD): SU transmits using $ \lbrace r_2 ,P_2 \rbrace$;
\item Channel is idle and detected busy (false alarm), denoted (FA): SU transmits using $ \lbrace r_1 ,P_1 \rbrace$;
\item Channel is idle and detected idle (correct detection), denoted (I-I): SU transmits using $ \lbrace r_2 ,P_2 \rbrace$.
\end{enumerate}
\vspace{-0.12 cm}
Approximating the PU interference term on the SU, $s_p(i)$, as an additional Gaussian noise, we can express the SU instantaneous channel
capacities in the above four scenarios as follows:
\begin{equation*}
C(i)_l=B\log (1+SNR_l z(i)), \;\;\;\; l=1, 2, 3, 4,
\end{equation*}
where $ SNR_1= \frac {P_1}{B(\sigma_n^2+\sigma^2_{s_p})} $, $ SNR_2= \frac {P_2}{B(\sigma_n^2+\sigma^2_{s_p})} $, $ SNR_3= \frac {P_1}{B(\sigma_n^2)} $ and $ SNR_4= \frac {P_2}{B(\sigma_n^2)} $.

We assume that the SU transmitter does not know the channel state information (CSI), to set the transmitted data rate every slot. Hence, $r_1$ and $r_2$ may be smaller or greater than the instantaneous channel capacity $C(i)$. Therefore, the channel can be either ON or OFF, depending on whether the fixed-transmission rate exceeds the instantaneous channel capacity or not. If the transmission rate is smaller than the instantaneous channel capacity the channel is said to be ON; otherwise, the channel is considered in the OFF state (outage state). When the channel is OFF, reliable communication is not attained, and hence, the information has to be resent. It is also assumed that a simple automatic repeat request (ARQ) mechanism is incorporated in the communication protocol to acknowledge the reception of data and to ensure that erroneous data is retransmitted. Accordingly, the effective transmission rate in the OFF states is zero. Thus, the developed Markov chain, models the sensing outcomes and the implications of its correctness/errors. Additionally, Markov chain, to be discussed later, captures the channel state (ON or OFF) as well.

The Markov chain is fully characterized by its transition probability matrix $\mathbf{R}_{M \times M}$ defined as:
\begin{equation*}\label{StateTr}
\mathbf{R}_{M \times M} =
\begin{bmatrix}
p_{i,j}
\end{bmatrix},  1 \leq i,j \leq M,
\end{equation*}
where $M$ is the number of states in the Markov chain. Along the same lines of \cite{akin2010effective}, the EC for such system model is expressed as follows:\footnote{The proof can be found in \cite[Ch.7]{chang2000performance}.}
\begin{equation}
EC(\theta)=\frac{\Lambda(-\theta)}{-\theta}=\max_{r1,r2} \frac{1}{-\theta}\log_esp(\mathbf{\Phi}(-\theta)\mathbf{R}) \label{EffectiveCapacity}
\end{equation}
The matrix $\mathbf{R}$ is the state transition matrix as defined above, where $sp(\mathbf{\Phi}(-\theta)\mathbf{R})$ is the spectral radius of the matrix $ \mathbf{\Phi}(-\theta) \mathbf{R}$ (i.e., the maximum of the absolute of all eigenvalues of the matrix).
Therefore, to reach a closed form expression for the EC, we need to get the eigenvalues of the matrix $\mathbf{\Phi}(-\theta) \mathbf{R}$. $ \mathbf{\Phi}(-\theta)$ is a diagonal matrix defined as $ \mathbf{\Phi}(-\theta) = diag(\phi_1(-\theta),\phi_2(-\theta),....,\phi_M(-\theta))$ whose diagonal elements are the moment generating functions of the Markov process in each of the $M$ states.

Next, we study and analyze the EC of the system under the interference model  for the no feedback system, where the PU feedback messages are not exploited at the SU side.

\subsection{No Feedback Exploitation (Baseline)}

\begin{figure}
\centering
\includegraphics[width=.78\linewidth,height=0.24\textheight]{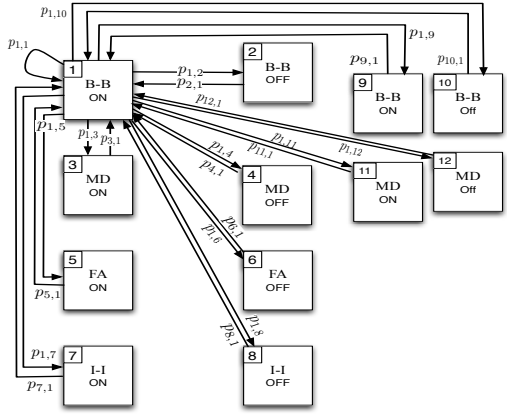}
\caption{The Markov chain model of the no feedback system.}\label{MarcovCHNoFB}
\vspace{-.4cm}
\end{figure}

First, we consider the case where no feedback is overheard from the PU receiver by the SU transmitter. In order to construct the Markov chain that models this system, we need to know how many states are required to model all the states of the system. We find that a 12-state Markov chain is required to model the system with no feedback. The Markov chain for the system is shown in Fig. \ref{MarcovCHNoFB} where only the transitions from and into the first state are shown for simplicity of presentation.

Therefore, the states from 1 to 8 model the system when the PU accesses the medium with prior probability $ \rho $ as mentioned before. We need 4 states to model the sensing process outcome, namely, (B-B, MD, FA and I-I). Moreover, to capture the ON/OFF channel states, we double the previously mentioned 4 states. The remaining 4 states are needed to capture the differences between feedback/No feedback systems and to properly compare between them. These 4 states will model the system when the PU receiver sends a NACK; in these states, the PU will access the channel with probability one since the PU will be retransmitting the erroneously received primary packet from the previous slot. Therefore, state 9 (ON) and state 10 (OFF) represent the (B-B) (i.e., correct sensing case), when the PU accesses the medium for a primary packet retransmission. Same interpretation applies to state 11 (ON) and state 12 (OFF), for the (MD) case (i.e., incorrect sensing result) when the PU accesses the channel for a primary packet retransmission. The probability to move into any of the last 4 states (9 to 12) will be a function of the primary link outage probability $\Pr(\rm outage)$.

The PU outage probability is the probability that the PU data rate is less than the channel instantaneous capacity, which is a function of the SU transmission power and noise power. In the underlying ``SINR interference model", it should be clear that there exists two different outage probabilities over the primary link $\Pr(\rm outage_1)$ and $\Pr(\rm outage_2) $ corresponding to the outage probabilities when the SU is transmitting with power levels $P_1$ and $P_2$, respectively. Since we assume that the only reason for failure over the primary link is due to the link outage, we will have $\Pr(\rm outage_1)=\Pr(NACK')$ and $\Pr(\rm outage_2)=\Pr(NACK'')$; $\Pr(NACK')$ is the probability of hearing a NACK message from the primary receiver at a given time slot assuming that the SU was transmitting with power $ P_1 $ and rate $ r_1 $ at that time slot. Similarly, $\Pr(NACK'')$ is the probability of hearing a NACK message at any time slot assuming that the SU was transmitting with power $ P_2 $ and rate $ r_2 $. Since $P_1 < P_2$, hence, the PU will suffer lower interference when the SU is transmitting with the lower power level $ P_1 $; therefore, it is evident that $\Pr(NACK') < \Pr(NACK'')$. In order to use same notations in this subsection and the next subsection, in which we present the feedback-aided system, we use $ \Pr(NACK)$ instead of $\Pr(\rm outage)$. 

In order to fully characterize our Markov chain model, the transition probabilities of our model are characterized as follows:
\begin{equation*}
\begin{split}
p&_{1,1} =\\ & \rho P_d \Pr (  r_1<C_1(i+TB)|r_1<C_1(i) )  (1-\Pr(NACK '))\\ &
=\rho P_d \Pr ( z(i+TB)>\alpha _1 |z(i)> \alpha _1  ) (1-\Pr(NACK ')),
\end{split}
\end{equation*}
where $ \alpha _1 = \frac {2^{\frac{r_1}{B}}}{SNR_1}$, the term $ \Pr (  r_1<C_1(i+TB)|r_1<C_1(i) ) $ represents the probability that the channel is ON (SU not in outage), $\rho$ is the prior probability of the primary channel being busy, $P_d$ is the probability of detection as defined in (\ref{prob_d}) and  $ (1-\Pr(NACK ')) $ is the probability of no outage in the primary link. In a block fading channel model, the fading changes independently from one block to another. Hence, $p_{1,1}$ can be further simplified as
\begin{equation}
\begin{split}
p_{1,1}&=  \rho P_d \Pr ( z(i+TB)> \alpha _1 ) (1-\Pr(NACK ')) \\ &= \rho P_d P ( z> \alpha _1 ) (1-\Pr(NACK ')).
\end{split}
\end{equation}
\begin{equation*}
p_{i,1} = \left\{ \begin{array}{rl}
  p_1=\rho P_d \Pr ( z > \alpha _1 ),  &\mbox{$i=5, 6, ..., 12$} \\
   p_1(1-\Pr(NACK')),&\mbox{$i=1,2$} \\
    p_1(1-\Pr(NACK'')),&\mbox{$i=3,4$}
       \end{array} \right.
\end{equation*}
\begin{equation*}
p_{i,2} = \left\{ \begin{array}{rl}
  p_2=\rho P_d \Pr ( z < \alpha _1 ),  &\mbox{$i=5, 6, ..., 12$} \\
   p_2(1-\Pr(NACK')),&\mbox{$i=1,2$} \\
    p_2(1-\Pr(NACK'')),&\mbox{$i=3,4$}
       \end{array} \right.
\end{equation*}
Similarly,
\begin{equation*}
p_{i,3}= p_3 = \rho (1-P_d) \Pr ( z > \alpha _2 ),   \;\;\;\;   \;\;\;\; i=5,6,..,12.
\end{equation*}
\noindent
where $ \alpha _2 = \frac {2^{\frac{r_2}{B}}}{SNR_2}$, $ \alpha _3 = \frac {2^{\frac{r_1}{B}}}{SNR_3}$ and $ \alpha _4 = \frac {2^{\frac{r_2}{B}}}{SNR_4}$.

As in states 1 and 2, other transition probabilities for states $4,5,\cdots,8$ can be characterized in the same way\footnote{Transition probabilities for states $4,5,\cdots,8$ are omitted due to space limitations}. However, for states 9, 10, 11 and 12, the transition probabilities are different since the probability that the system enters these states is a function of the PU outage probability.
\begin{equation}
p_{i,9} = \left\{ \begin{array}{ll} \label{eqn_9}
  P_d \Pr(NACK')  \Pr ( z > \alpha _1 ),  &\mbox{$i=1,2$} \\
   P_d \Pr(NACK'')  \Pr ( z > \alpha _1 ),&\mbox{$i=3,4$}  \\
0, &\mbox{otherwise.}
       \end{array} \right.
\end{equation}
\begin{equation}
p_{i,11} = \left\{ \begin{array}{ll} \label{eqn_10}
  (1-P_d) \Pr(NACK')  \Pr ( z > \alpha _2 ),  &\mbox{$i=1,2$} \\
   (1-P_d) \Pr(NACK'')  \Pr ( z > \alpha _2 ),&\mbox{$i=3,4$}  \\
0, &\mbox{otherwise.}
       \end{array} \right.
\end{equation}
Similar to (\ref{eqn_9}) and (\ref{eqn_10}), we characterize $p_{i,10}$ and $ p_{i,12} $ by only changing the probability of ON channel to OFF channel probability.

It is worth noting that, we assume that the system will never stay in states 9, 10, 11 or 12 for 2 successive time frames. This assumption implies that the PU will retransmit a packet one time. Despite the fact that this assumption is not practical, we adopt it to avoid packets accumulation in the primary queue. Moreover, this assumption does not affect our main contribution since it is fairly adopted to both systems. It is also clear that no transitions are permitted from states 5 up to 8 to states 9 up to 12 since it is impossible for the PU to get a NACK while being idle.

Accordingly, we have completely specified the transition probability matrix $\mathbf{R}_{12 \times 12}$. The moment generating function corresponding to each state depends on the effective rate of each state. Hence, $\mathbf{\Phi}(-\theta) = diag\lbrace e^{-(T-N) \theta r_1}, 1, e^{-(T-N) \theta r_2}, 1, e^{-(T-N) \theta r_1}, 1,\\ e^{-(T-N) \theta r_2}, 1, e^{-(T-N) \theta r_1}, 1, e^{-(T-N) \theta r_2}, 1   \rbrace $.

Since we will compare in the next section the average power transmitted by the SU under both systems. Thus, for this system, it is worth noting that the average power transmitted by the SU $ P_{avg_N} $ can be computed as follows:
\begin{equation}
P_{avg_N}=P_1 \sum _{i=1,2,5,6,9,10} \bar{p_i} + P_2 \sum _{i=3,4,7,8,11,12} \bar{p_i}.  \label{power_eqn1}
\end{equation}
where $\bar{p_i}$ is the steady state probability of state $i$ and $P_1$ and $P_2$ are the SU transmission powers as defined above.

\subsection{Feedback Aided Scheme}

In case the SU overhears, and exploits, the PU feedback for channel access two scenarios arise. First, if the SU overhears a NACK, it does not sense the channel in the next time slot as the PU will retransmit with probability one. Hence, the SU transmits with low power and rate (i.e. $P_1$ and $r_1$) in this slot. On the other hand, if an ACK or no FB is overheard by the SU, it behaves normally (similar to the baseline system) and starts sensing the channel at the beginning of the next time slot, since the PU may/may not be active.

\begin{figure}
\centering
\includegraphics[width=.78\linewidth,height=0.24\textheight]{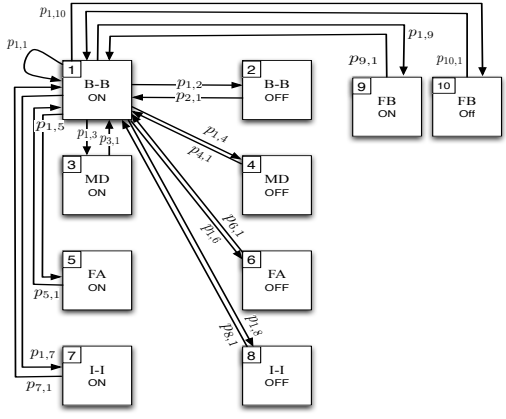}
\caption{The Markov chain model of the feedback-aided system.} \label{MarcovCHFB}
\vspace{-.4cm}
\end{figure}

In this system, we have a 10-state Markov chain, as shown in Fig. \ref{MarcovCHFB}. States 1 through 8 model exactly the same behaviour as the baseline no feedback system. On the other hand, state 9 (ON) and state 10 (OFF) model the event, pointed out above, which captures the fundamental difference between the two systems, that is, the secondary transmitter overhearing a NACK message from the primary receiver.
Typically, overhearing the PU feedback would only affect the last four states which would be reduced to two states corresponding to hearing a primary receiver NACK as shown in Fig. \ref{MarcovCHFB}. Hence, the SU would transmit with power $P_1$  and rate $r_1$.

Note that the state transition probabilities between the first 8 states are the same as the no feedback system. This is attributed to the fact that both systems experience exactly the same behaviour and dynamics in case of ACK or no feedback. On the other hand, the transition probabilities for states 9 and 10 are as follows:
\begin{equation}
p_{i,9} = \left\{ \begin{array}{ll} \label{eqn_11}
  \Pr(NACK')  \Pr ( z > \alpha _1 ),  &\mbox{$i=1,2$} \\
  \Pr(NACK'')  \Pr ( z > \alpha _1 ), &\mbox{$i=3,4$}  \\
0, &\mbox{otherwise,}
       \end{array} \right.
\end{equation}
\noindent
where $ \Pr(NACK') $ is the probability of hearing a NACK from the primary receiver when the SU transmits using $ P_1 $ and $ r_1 $ which is the case at state 1 and state 2. However, $ \Pr(NACK'') $ is the probability of hearing a NACK from the primary receiver when the SU transmits using $ P_2 $ and $ r_2 $ at state 3 and state 4. On the other hand, being in any other state can never cause the SU to overhear a NACK message from the PU, hence, no transitions to state 9. Similarly, we characterize $p_{i,10}$ as in (\ref{eqn_11}) by changing the probability of the ON channel to the OFF channel probability.

Accordingly, we have completely specified the transition probability matrix $\mathbf{R}_{12 \times 12}$. Since we characterize the EC via the spectral radius of matrix $\mathbf{\Phi}(-\theta)\mathbf{R}$, the moment generating functions corresponding to each state depends on the effective rate of that state, where
\begin{equation*}
\begin{split}
\mathbf{\Phi}(-\theta) &= diag\lbrace e^{-(T-N) \theta r_1}, 1, e^{-(T-N) \theta r_2} \\& , 1, e^{-(T-N) \theta r_1}, 1, e^{-(T-N) \theta r_2}, 1, e^{-T \theta r_1}, 1  \rbrace.
\end{split}
\end{equation*}

Next, the main result of the paper is presented in the following theorem (the proof of the theorem is given in the Appendix).

\noindent\textbf{Theorem 1:} { \it Under the SINR interference model, the effective capacity of the feedback-aided system is always greater than the effective capacity of the no feedback system}.

The average power transmitted by the SU under the feedback system is given by
\begin{equation} 
P_{avg_F}=P_1 \sum _{i=1,2,5,6,9,10} \tilde{p_i} + P_2 \sum _{i=3,4,7,8} \tilde{p_i}. \label{power_eqn2}
\end{equation}
where $\tilde{p_i}$ is the steady state probability of state $i$ in the feedback system and $P_1$ and $P_2$ are the SU transmission powers as defined above.

Simulation results presented next section support our main analytical result established in Theorem 1 and provides further insights about the key parameters of the system.

\section{Numerical Results}

\begin{figure}
\centering
\includegraphics[width=1\linewidth,height=0.18\textheight]{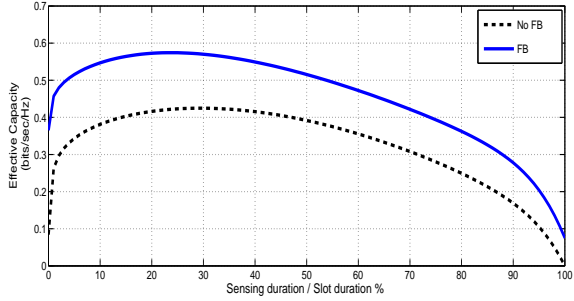}
\caption{Effective capacity as a function of the sensing duration $ N $, percentage of the slot duration}\label{EC_N}
\vspace{-.4cm}
\end{figure}

\begin{figure}
\centering
\includegraphics[width=1\linewidth,height=0.18\textheight]{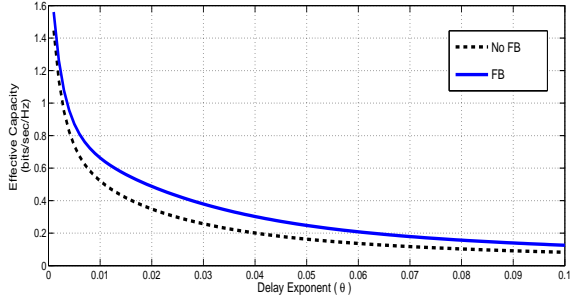}
\caption{Feedback-aided system outperforms the No feedback system under the interference model.}\label{fig_EC}
\vspace{-.4cm}
\end{figure}

%

In this section, we present numerical results that emphasize our contributions. The system parameters numerical values are: $SNR_1 = 6.9 $ db , $SNR_2 = 10$ db, $SNR_3 = 30.7$ db, $SNR_4 = 40$ db, $\Pr(NACK') = 0.3$, $\Pr(NACK'') = 0.9$, $r_1=1000 $ bps, $r_2=10000 $ bps and $\rho = 0.7$. We also set $T=0.1$ sec, $N=0.026$ sec, $ \lambda=1.7 $ and $B=1000$ Hz. Note that the values for $r_1$ and $r_2$ are obtained by simple numerical search.

In Fig. \ref{EC_N}, the EC is plotted versus the sensing duration $N$, represented as a percentage of the time slot duration $T$. It shows that, for a fixed delay exponent $ \theta $ and other system parameters, there is an optimum sensing duration $ N $ that maximizes the SU effective capacity, (nearly at $ N/T= 26 \% $). The maximum observed, for both systems, suggests that there is a trade-off which we explain here. First, for small sensing periods, sensing errors are more frequent, and hence, interference is higher and more packets are lost, therefore, EC becomes lower. On the other hand, for longer sensing periods, sensing becomes more reliable, yet, the portion of the time slot left for data transmission, i.e. $ T-N $ becomes smaller, yielding lower effective capacity. This, in turn, gives rise to the quest for optimal sensing duration that strikes a balance for the aforementioned fundamental trade-off which we shed some light on in this paper. Moreover, we show that the no feedback (baseline) system needs to sense $ 26\% $ of the slot duration in order to reach a maximum EC. While the feedback-aided system needs to sense less than $1 \%$, to give the maximum EC of the no feedback system.

Fig. \ref{fig_EC} plots the EC for the feedback-aided and the no feedback systems versus the delay exponent $\theta$ for a sensing duration of $N=0.026$ for which $ P_f=0.0012 $ and $ P_d=0.7705 $ according to (\ref{prob_f}) and (\ref{prob_d}). Clearly, as the delay exponent $\theta$ increases (stricter delay requirements), the effective capacity (the maximum rate that the network can support in bit/sec/hertz)  decreases. The same result can be easily expected from the  EC deﬁnition in (\ref{eqn_EC}). Moreover, it is shown that feedback exploitation yields secondary user EC gains. As $ \theta $ increases the performance gain decreases since stricter QoS constraint limits the secondary user throughput, hence, both systems converge to the maximum arrival rate that can be supported by the secondary link \cite{wu2003effective}. Exploiting the feedback yields an EC improvement of 36$ \% $  at $\theta=0.02$. 
%

\begin{figure}
\centering
\includegraphics[width=1\linewidth,height=0.18\textheight]{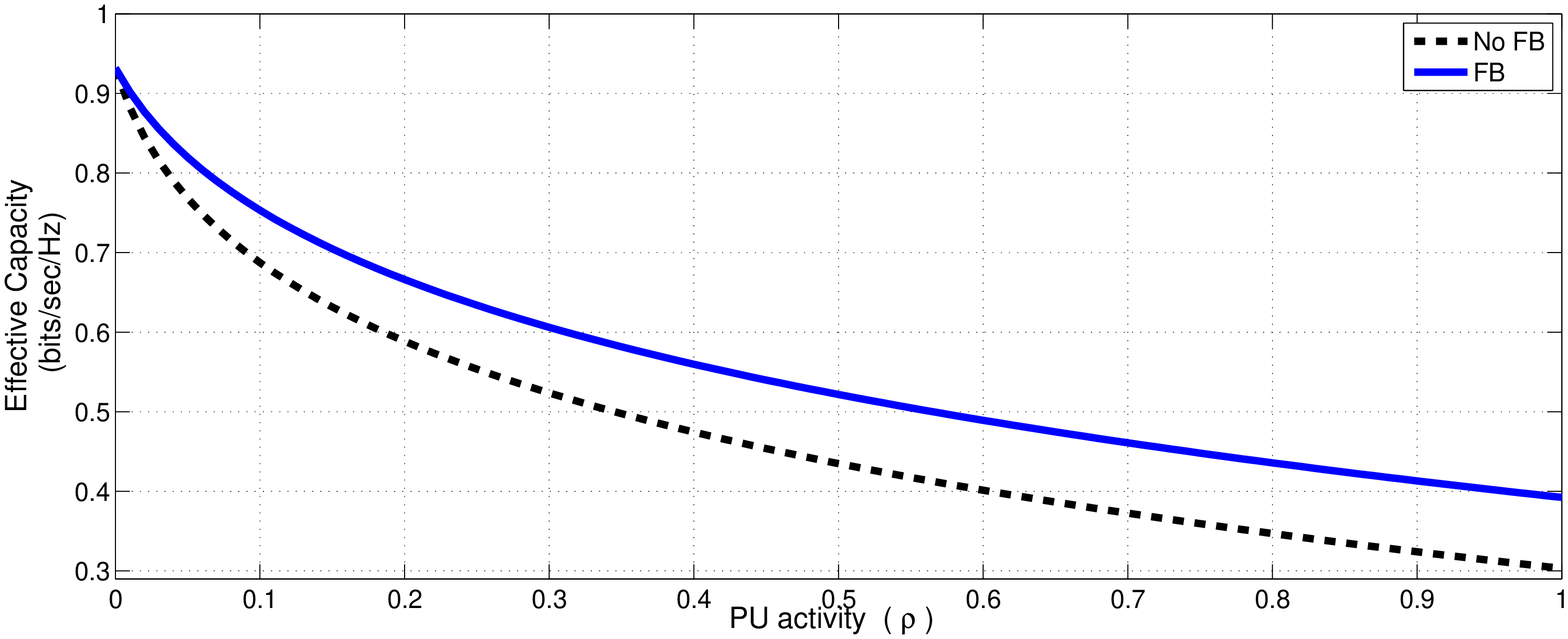}
\caption{Effective Capacity Versus the PU active prior probability $ \rho $}\label{EC_rho}
\vspace{-.4cm}
\end{figure}

In Fig. \ref{EC_rho}, the EC (bit/sec/Hz) is plotted versus the PU access probability $ \rho $. It is shown that the EC of both systems decreases as the PU use the channel more frequently. Moreover, it is clear that the amount of improvement due to exploiting the PU feedback increases as $ \rho $ increases. As $ \rho $ increases, the PU is occupying the medium more frequently, and hence, sends more packets, receives more feedback messages and the SU senses the channel busy in more slots. Therefore, EC gains attributed to feedback overhearing increase. It is worth noting that, even at PU prior probability $ \rho = 1$, the EC does not completely vanish. This is due to the interference model we adopt in this paper which allows the PU and SU to coexist as long as the interference is within tolerable levels.

\begin{figure}
\centering
\includegraphics[width=1\linewidth,height=0.18\textheight]{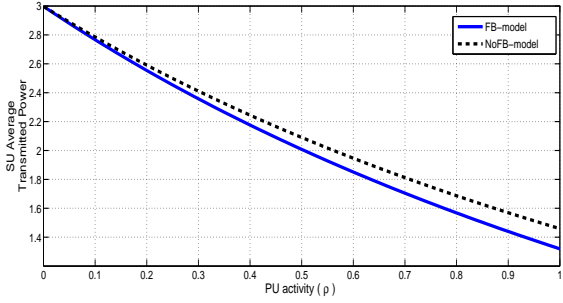}
\caption{Average transmitted power Versus the PU active prior probability $ \rho $, $ P_1 = 1 $ and $ P_2 = 2 $ unit power}\label{power_rho}
\vspace{-.4cm}
\end{figure}

In Fig. \ref{power_rho}, we plot the average transmit power, for both systems, versus the PU prior activity probability, $ \rho $. The average transmit power decreases as the PU activity ($ \rho $) increases, since the channel becomes busy with higher probability. The SU saves power in case of the feedback-aided scheme by avoiding outage and interference with the PU if a NACK is received, as derived in equations (\ref{power_eqn1}) and (\ref{power_eqn2}).  Assuming the system power parameters are given by $ P_1 =1 $ unit power and $ P_2 = 3 $ unit power, it is shown in Fig. \ref{power_rho} that for the feedback-aided system, the SU has a slightly lower average transmitted power..

\section{Conclusion and Future Work}\label{Concl}

In this paper, we analyze the effective capacity (EC) for a CR network with a SU that wishes to coexist with a PU sharing the same wireless medium. We proved analytically that exploiting the primary feedback overheard at the secondary transmitter yields performance gains and power savings for the SU EC, under the interference model. In formulating our problem, we assume that the SU eavesdrops on the primary link's feedback without losing any resources, i.e., the SU transmits at lower power, without sensing, upon hearing a NACK. Otherwise, the power level is determined based on the sensing results. This leads to significant gains in SU EC. In addition, we showed that feedback exploitation slightly reduces the SU average transmitted power.

For future work, the case of having more than one secondary link can be investigated, considering both cooperative and non-cooperative secondary networks. Also, investigating the feedback overhearing usefulness if a ``Collision model'' is adopted instead of the ``SINR Interference model''.

\section*{Appendix}
First, we quantify the EC for the no feedback (baseline) system and then extend the analysis to the feedback-aided system. The effective capacity of the system is governed by the spectral radius of the matrix $\mathbf{\Phi}_N(-\theta)\mathbf{R}_N$\footnote{The subscript $N$ refers to no feedback.}. From Linear Algebra, the spectral radius of a matrix is the maximum absolute eigenvalue.
\begin{proof}
\begin{equation*}
\small
\mathbf{\Phi}_N(-\theta)\mathbf{R}_N=\begin{bmatrix}
a_1k_1\mathbf{v}  &k_1 p_{1,9}  &k_1 p_{1,10}  &k_1 p_{1,11}  &k_1 p_{1,12} \\
a_1\mathbf{v}  &p_{2,9}  &p_{2,10}  &p_{2,11}  &p_{2,12} \\
a_2k_2 \mathbf{v}  &k_2p_{3,9}  &k_2p_{3,10}  &k_2p_{3,11}  &k_2p_{3,12} \\
a_2\mathbf{v}  &p_{4,9}  &p_{4,10}  &p_{4,11}  &p_{4,12} \\
k_1\mathbf{v}  &0  &0  &0  &0 \\
\mathbf{v}  &0  &0  &0  &0 \\
k_2\mathbf{v}  &0  &0  &0  &0 \\
\mathbf{v}  &0  &0  &0  &0 \\
k_1\mathbf{v}  &0  &0  &0  &0 \\
\mathbf{v}  &0  &0  &0  &0 \\
k_2\mathbf{v}  &0  &0  &0  &0 \\
\mathbf{v}  &0  &0  &0  &0
\end{bmatrix}
\end{equation*}
\noindent
where $ \mathbf{v} =\begin{bmatrix}
p_1 &p_ 2 &p_ 3 &p_ 4 &p_ 5 &p_6 &p_7  &p_8
\end{bmatrix}$,
\vspace{-0.005 cm}
$a_1~=~ 1-\Pr(NACK'), \;\;a_2= 1-\Pr(NACK''), \;\;k_1= e^(-(T-N)r_1\theta) \;\;$ and $ \;\;  k_2= e^(-(T-N)r_2\theta) $. From linear algebra, the eigenvalue basic equation is
\begin{equation}
\mathbf{w}\mathbf{\Phi}_N(-\theta)\mathbf{R}_N=\lambda _N\mathbf{w}\label{matrix},
\end{equation}
\noindent
where $ \mathbf{w}= \begin{bmatrix}
w_1 &w_2  &w_3  & \cdots   &w_{12}
\end{bmatrix} $ is the eigenvector corresponding to the eigenvalue $ \lambda _N $ of the matrix $ \mathbf{\Phi}_N(-\theta)\mathbf{R}_N $. Performing matrix multiplication in (\ref{matrix}), we obtain
\begin{equation*}
\begin{split}
 &a_1k_1w_1p_1+a_1w_2p_1+a_2k_2w_3p_1+a_2w_4p_1+k_1w_5p_1+w_6p_1\\&+k_2w_7p_1+w_8p_1+k_1w_9p_1+w_{10}p_1+k_2w_{11}p_1+w_{12}p_1
 \\&=\lambda _N w_1. 
\end{split}
\end{equation*}
Or equivalently $c_1p_1=\lambda _N w_1,
$
where
\begin{equation*}
\begin{split}
c_1=&a_1k_1w_1+a_1w_2+a_2k_2w_3+a_2w_4+k_1w_5+w_6\\&+k_2w_7+w_8+k_1w_9+w_{10}+k_2w_{11}+w_{12}.
\end{split}
\end{equation*}
Similarly, it can be shown that
$
\;\;\;\; c_1p_2=\lambda_N w_2.
$

In general,
\begin{equation}\label{p2}
c_1p_m=\lambda _N w_m \;\;\;\; m=3, 4, ...,8.
\end{equation}
\begin{equation*}
k_1p_{1,9}w_1+p_{2,9}w_2+k_2p_{3,9}w_3+p_{4,9}w_4=\lambda _N w_9.
\end{equation*}
\begin{equation*}
w_9=\frac{k_1p_{1,9}w_1+p_{2,9}w_2+k_2p_{3,9}w_3+p_{4,9}w_4}{\lambda _N}.
\end{equation*}
\begin{equation*}
w_{10}=\frac{k_1p_{1,10}w_1+p_{2,10}w_2+k_2p_{3,10}w_3+p_{4,10}w_4}{\lambda _N}.
\end{equation*}
\begin{equation*}
w_{11}=\frac{k_1p_{1,11}w_1+p_{2,11}w_2+k_2p_{3,11}w_3+p_{4,11}w_4}{\lambda _N}.
\end{equation*}
\begin{equation*}
w_{12}=\frac{k_1p_{1,12}w_1+p_{2,12}w_2+k_2p_{3,12}w_3+p_{4,12}w_4}{\lambda _N}
\end{equation*}
\vspace{-0.2 cm}
Substituting for $ w_{9}, w_{10}, w_{11}$ and $w_{12}$ into $c_1$, we get
\begin{equation*}\small
\begin{split}
c_1=&\left( a_1k_1+\frac{k_1^2p_{1,9}+k_1p_{1,10}+k_1k_2p_{1,11}+k_1p_{1,12}}{\lambda _N}\right)w_1 \\&+\left( a_1+\frac{k_1p_{2,9}+p_{2,10}+k_2p_{2,11}+p_{2,12}}{\lambda _N}\right)w_2\\ &+\left( a_2k_2+\frac{k_1k_2p_{3,9}+k_2p_{3,10}+k_2^2p_{3,11}+k_2p_{3,12}}{\lambda _N}\right)w_3\\&+\left( a_2+\frac{k_1p_{4,9}+p_{4,10}+k_2p_{4,11}+p_{4,12}}{\lambda _N}\right)w_4\\&+k_1w_5+w_6+k_2w_7+w_8.
\end{split}
\end{equation*}
\vspace{-0.2 cm}
Forming a linear combination from equation (\ref{p2}) as follows:
\begin{equation*}\small
\begin{split}
&\left( a_1k_1+\frac{k_1^2p_{1,9}+k_1p_{1,10}+k_1k_2p_{1,11}+k_1p_{1,12}}{\lambda _N}\right) \times \lambda _N w_1\\&+ \left( a_1+\frac{k_1p_{2,9}+p_{2,10}+k_2p_{2,11}+p_{2,12}}{\lambda _N}\right) \times \lambda _N w_2\\&+
\left( a_2k_2+\frac{k_1k_2p_{3,9}+k_2p_{3,10}+k_2^2p_{3,11}+k_2p_{3,12}}{\lambda _N}\right) \times \lambda _N w_3\\
&+\left( a_2+\frac{k_1p_{4,9}+p_{4,10}+k_2p_{4,11}+p_{4,12}}{\lambda _N}\right) \times \lambda _N w_4\\ 
\end{split}
\end{equation*}
\begin{equation*}\small
\begin{split}
&+ k_1 \times \lambda _N w_5+\lambda _N w_6+k_2\times \lambda _N w_7+\lambda _N w_8.
\end{split}
\end{equation*}
\noindent
This results in
\begin{equation*}\small
\begin{split} 
&c_1\bigg[\left( a_1k_1+\frac{k_1^2p_{1,9}+k_1p_{1,10}+k_1k_2p_{1,11}+k_1p_{1,12}}{\lambda _N}\right)  p_1 \\&+
\left( a_1+\frac{k_1p_{2,9}+p_{2,10}+k_2p_{2,11}+p_{2,12}}{\lambda _N}\right) p_2 \\&+
\left( a_2k_2+\frac{k_1k_2p_{3,9}+k_2p_{3,10}+k_2^2p_{3,11}+k_2p_{3,12}}{\lambda _N}\right)p_3\\&+
\left( a_2+\frac{k_1p_{4,9}+p_{4,10}+k_2p_{4,11}+p_{4,12}}{\lambda _N}\right) p_4\\&+k_1p_5+p_6+k_2p_7+p_8 \bigg] = c_1 \lambda _N.
\end{split}
\end{equation*}
For non-negative matrices, the spectral radius as well as the eigenvector corresponding to it are positive according to the Perron-Frobenius Theorem \cite{frobenius1912matrizen}, which means that $c_1 \neq 0$ for the spectral radius, and hence, it can be canceled from the two sides of the last equation.
Forming a second order polynomial in $\lambda _N$ as follows:
\begin{equation}
\lambda _N ^{2} - a' \lambda _N - b' = 0. \label{eqn_lamd_N}
\end{equation}
\noindent
It is clear that, $a'$ and $b'$ are in terms of $a_1$, $a_2$, $k_1$, $k_2$ and $p_1 \dots p_8 $, however, for now we focus only on $b'$.
\begin{equation*}
\begin{split}
b'=&(k_1^2p_{1,9}+k_1p_{1,10}+k_1k_2p_{1,11}+k_1p_{1,12})p_1\\&+(k_1p_{2,9}+p_{2,10}+k_2p_{2,11}+p_{2,12})p_2\\&+(k_1k_2p_{3,9}+k_2p_{3,10}+k_2^2p_{3,11}+k_2p_{3,12})p_3\\&+(k_1p_{4,9}+p_{4,10}+k_2p_{4,11}+p_{4,12})p_4.
\end{split}
\end{equation*}

In order to complete the proof, we now shift our attention to the feedback-aided system. Characterizing the EC through the eigenvalue of matrix $ \mathbf{\Phi}_F(-\theta)\mathbf{R}_F $\footnote{The subscript $F$ refers to feedback.} Along the same lines of the no feedback case,
\begin{equation*}
\mathbf{\hat{w}}\mathbf{\Phi}_F(-\theta)\mathbf{R}_F= \lambda_F\mathbf{\hat{w}}.
\end{equation*}
\begin{equation*}
\mathbf{\hat{w}}= \begin{bmatrix}
\hat{w}_1 &\hat{w}_2  &\hat{w}_3  & ...   &\hat{w}_{10}  \label{matrix2}
\end{bmatrix}.
\end{equation*}
\begin{equation*}
\mathbf{\Phi}_F(-\theta)\mathbf{R}_F=\begin{bmatrix}
a_1k_1\mathbf{v} &k_1 \hat{p}_{1,9}  &k_1 \hat{p}_{1,10}   \\
a_1\mathbf{v} &\hat{p}_{2,9}  &\hat{p}_{2,10}  \\
a_2k_2 \mathbf{v} &k_2\hat{p}_{3,9}  &k_2\hat{p}_{3,10}   \\
a_2\mathbf{v} &\hat{p}_{4,9}  &\hat{p}_{4,10}  \\
k_1\mathbf{v} &0  &0  \\
\mathbf{v} &0  &0   \\
k_2\mathbf{v} &0  &0  \\
\mathbf{v} &0  &0   \\
k_1\mathbf{v} &0  &0   \\
\mathbf{v} &0  &0  \\
k_2\mathbf{v} &0  &0  \\
\mathbf{v} &0  &0
\end{bmatrix}.
\end{equation*}
Similarly, we write the matrix multiplication output as follows:
\begin{equation*}
\hat{c}_1\hat{p}_m=\lambda _F \hat{w}_m \;\;\;\; m=1, 2, ..., 8.
\end{equation*}
\begin{equation*}
\hat{w}_9=\frac{k_1\hat{p}_{1,9}\hat{w}_1+\hat{p}_{2,9}\hat{w}_2+k_2\hat{p}_{3,9}\hat{w}_3+\hat{p}_{4,9}\hat{w}_4}{\lambda _F}.
\end{equation*}
\begin{equation*}
\hat{w}_{10}=\frac{k_1\hat{p}_{1,10}\hat{w}_1+\hat{p}_{2,10}\hat{w}_2+k_2\hat{p}_{3,10}\hat{w}_3+\hat{p}_{4,10}\hat{w}_4}{\lambda _F}.
\end{equation*}
\begin{equation*}
\begin{split}
\hat{c}_1&=\left(  a_1k_1+\frac{k_1^2\hat{p}_{1,9}+k_1\hat{p}_{1,10}}{\lambda _F}\right) \hat{w}_1\\&+\left(  a_1+\frac{k_1\hat{p}_{2,9}+\hat{p}_{2,10}+}{\lambda _F}\right)\hat{w}_2\\&+ \left( a_2k_2+\frac{k_1k_2\hat{p}_{3,9}+k_2\hat{p}_{3,10}}{\lambda _F}\right)\hat{w}_3 \\& +\left(  a_2+\frac{k_1\hat{p}_{4,9}+\hat{p}_{4,10}}{\lambda _F}\right)\hat{w}_4+k_1\hat{w}_5+\hat{w}_6+k_2\hat{w}_7+\hat{w}_8
\end{split}
\end{equation*}
Following the same steps in the no feedback system, we construct a quadratic equation in $ \lambda _F $:
\begin{equation}
\lambda_F ^{2} - a'' \lambda_F - b'' = 0. \label{eqn_lamd_F}
\end{equation}
\noindent
\begin{equation*}
\begin{split}
b''=&(k_1^2\hat{p}_{1,9}+k_1\hat{p}_{1,10})p_1+(k_1\hat{p}_{2,9}+\hat{p}_{2,10})p_2\\&+(k_1k_2\hat{p}_{3,9}+k_2\hat{p}_{3,10})p_3+(k_1\hat{p}_{4,9}+\hat{p}_{4,10})p_4.
\end{split}
\end{equation*}

Formulating $a', a'', b'$ and $b''$ in terms of the non-negative quantities $a_1, a_2, k_1, k_2 $ and transition probabilities, we found that, $a'=a'' > 0$. While, $b' > b''$, which directly implies from (\ref{eqn_lamd_N}) and (\ref{eqn_lamd_F}) that $\lambda _N > \lambda _F$ (i.e. the spectral radius of the feedback-aided system is lower and hence gives higher effective capacity), since the EC is proportional to $-\log{(spectral \; radius)}$ as given in (\ref{EffectiveCapacity}).
\end{proof}

\vspace{-0.2 cm}
\bibliographystyle{IEEEtran}
\bibliography{IEEEabrv,ddd}

\end{document}